\documentclass[bg]{copernicus} 

\usepackage{graphicx}
\usepackage{pdfpages}
\begin{document}

\nolinenumbers

\title{Bioclimatic change as a function of global warming from CMIP6 climate projections}


\Author[1]{Morgan}{Sparey}
\Author[1]{Peter}{Cox}
\Author[1,2]{Mark S.}{Williamson}

\affil[1]{College of Engineering, Mathematics and Physical Sciences, University of Exeter, Exeter, EX4 4QF, UK }
\affil[2]{Global Systems Institute, University of Exeter, Exeter, EX4 4QE, UK}




\correspondence{Morgan Sparey (ms959@exeter.ac.uk)}

\runningtitle{Bioclimatic change as a function of global warming from CMIP6 climate projections}

\runningauthor{Sparey et al}

\received{}
\pubdiscuss{} 
\revised{}
\accepted{}
\published{}


\firstpage{1}

\maketitle

\begin{abstract}
Climate change is predicted to lead to major changes in terrestrial ecosystems \citep{1}. However, significant differences in climate model projections for given scenarios of greenhouse gas emissions \citep{2}, continue to hinder detailed assessment. Here we show, using a traditional Köppen-Geiger bioclimate classification system \citep{3}, that the latest CMIP6 Earth System Models actually agree very well on the fraction of the global land-surface that will undergo a significant change per degree of global warming. Data from ‘historical’ and ‘ssp585’ model runs are used to create bioclimate maps at various degrees of global warming, and to investigate the performance of the ensemble mean when classifying climate data into discrete categories. Using a streamlined Köppen-Geiger scheme with 13 classifications, global bioclimate classification maps at 2K and 4K of global warming above a 1901 - 1931 reference period are presented. These projections show large shifts in bioclimate distribution, with an almost exclusive change from colder, wetter bioclimates to hotter, dryer ones.  Historical model run performance is assessed and examined by comparison with the bioclimatic classifications derived from the observed climate over the same time period. The fraction ($f$) of the land experiencing a change in its bioclimatic class as a function of global warming ($\Delta T$) is estimated by combining the results from the individual models. Despite the discrete nature of the bioclimatic classification scheme, we find only a weakly-saturating dependence of this fraction on global warming $f = 1-e^{-0.17 \Delta T}$, which implies about 12\% of land experiencing a significant change in climate, per 1K increase in global mean temperature between the global warming levels of 1 and 3K. Therefore, we estimate that stabilising the climate at 1.5K rather than 2K of global warming, would save over 7 million square kilometres of land from a major bioclimatic change.
\end{abstract}


\introduction  
Understanding the impacts that climate change will have at a regional level yields vital information for adaptation to climate change. Furthermore, quantifying the performance of climate models is important for the continued improvement of climate models, and for understanding the regional areas where particular models under perform. Here we use the Köppen-Geiger (KG) bioclimate classification to examine and quantify changes in biome under various levels of projected future global warming within the Coupled Model Intercomparison Project phase 6 (CMIP6) climate models. We examine the differences between climate projections in terms of traditional KG classifications which summarise the aspects of regional climates which are most relevant to biomes, and therefore to the impacts of climate change on the natural environment. To remove the zeroth order uncertainty that arises from different climate model sensitivities to radiative forcing \citep{4,5}, and to make our results relevant to the Paris climate targets, we look specifically at changes in KG classification at different levels of global warming (1.5K, 2K, and 4K). A streamlined KG scheme is also implemented to visually demonstrate the impacts of warming on global biome distribution.

Climate classification empirically separates regions of the globe based on climate data. The KG bioclimate classification scheme is one of the most established, first developed by Wladimir Koppen \citep{3} and then enhanced by Rudolf Geiger. The original KG classification scheme consists of thirty separate bioclimates. These classifications are based on monthly average temperature and precipitation at each location. The seasonality of these variables, combined with threshold values, determines the bioclimate classification of the region \citep{6}.

CMIP6 is an international collaboration to run a standardised set of potential future scenarios with a range of climate models developed at various institutions. The results make a compelling case for the need to further prioritise climate change mitigation policies. However, this may not be immediately clear to public or policy makers. Improved understanding of the consequences of climate change is needed,  and climate classification schemes can help in that respect.

Bioclimate classification systems, such as the KG and Holdridge schemes \citep{8}, have been used to map regions or even the entire globe. These maps have been created using observational \citep{9} as well as climate model data, the latter including CMIP5 \citep{10} and CMIP6 climate models \citep{11}. Despite the changes and updates suggested by various authors, the classification scheme as originally developed by Köppen, and updated by Geiger is still the most popular climate classification system. The KG system has been applied to a broad spectrum of scientific interests, including to locally adjust an irradiation model \citep{12}, in hydrological studies \citep{13}, and in modelling the distribution of Lyme disease \citep{14}. Although bioclimate maps for specific years (such as 2100) have previously been created \citep{15}, an area that is less explored are global KG climate maps at specific levels of global warming.

Here we present KG classification maps at 1.5K, 2K, and 4K of global warming above reference period levels (taken as the 1901 – 1931 global mean temperature). Due to the 30 different classifications in the traditional KG scheme, it can be difficult to identify the changes in bioclimate classification, we present a novel “streamlined” classification system that allows for easy identification of bioclimate change, with a naming scheme that is more intuitive. To enhance this, classification change matrices have been produced to quantify the changes displayed in the maps.

Furthermore, we utilise the KG system as an exploratory technique for understanding CMIP6 model output, the KG classification scheme has been previously applied to CMIP5 data to to evaluate simulations \citep{19}. By comparing the classifications given by models under the historical experimental run to the known historical observational values, and by assessing model deviation from their initial classifications we gain insight into the performance and behaviours of individual models as well as their ensemble mean.

\section{Methods}
\subsection{Köppen-Geiger classification scheme}
The Köppen-Geiger (KG) classification scheme has been described extensively in other publications \citep{6,15}. The scheme has also undergone many alterations. Here we follow \citep{6}, whose criteria for each classification are given in Table 1.
\begin{table*}[h]
\begin{tabular}{llllllll}
\multicolumn{3}{l}{\textbf{Classification}} & \textbf{Criteria} & \multicolumn{3}{l}{\textbf{Classification}} & \textbf{Criteria} \\
A &  &  & Tmin ≥ 18˚C & D &  &  & Tmin \textless 0˚C, Tmax ≥ 10˚C \\
 & F &  & Pmin ≥ 6 cm/month &  & W &  & Pswet ≥ 10*Pwdry \\
 & S &  & Pmin ≥ 100-(Pyear*10/25) &  & S &  & 3*Psdry \textless Pwwet \\
 & W &  & Pmin \textless 100-(Pyear*10/25) &  & F &  & Neither W nor S \\
B &  &  & Pyear*10 \textless 10*Pthresh &  &  & a & Tmax ≥ 22˚C, Months above 10˚C ≥ 4 \\
 & W &  & Pyear*10 \textless 5*Pthresh &  &  & b & Tmax \textless 22˚C, Months above 10˚C ≥   4 \\
 & S &  & Pyear*10 ≥ 5*Pthresh &  &  & c & 0 \textless Months above 10˚C \textless 4, not   A or B or D \\
 &  & h & Tavg ≥ 18˚C &  &  & d & Tmin \textless -38˚C, 0 \textless Months above   10˚C \textless 4 \\
 &  & k & Tavg \textless 18˚C & E &  &  & Tmax \textless 10˚C \\
C &  &  & 0˚C ≤ Tmin \textless 18˚C, Tmax ≥ 10˚C &  & T &  & 0˚C ≤ Tmax \textless 10˚C \\
 & W &  & Pwdry \textless Pswet/10 &  & F &  & 0˚C ≥ Tmax \\
 & S &  & Pwwet ≥ 3*Psdry, Psdry \textless 4 &  &  &  &  \\
 & F &  & Neither W nor S &  &  &  &  \\
 &  & a & Tmax ≥ 22˚C, Months above 10˚C ≥ 4 &  &  &  &  \\
 &  & b & Tmax \textless 22˚C, Months above 10˚C ≥   4 &  &  &  &  \\
 &  & c & 0 \textless Months above 10˚C \textless 4, not   A or B &  &  &  &
\end{tabular}
\caption{\label{tab:table-name} Classification criteria for the Köppen-Geiger classification scheme. Tmin = Average temperature of month with lowest average temperature. Tmax = Average temperature of month with highest average temperature. Pmin = Average precipitation of driest month. Pmax = Average precipitation of wettest month. Tavg = Mean annual temperature. Pyear = Mean annual precipitation. Pthresh varies according to the following rules (if 70\% of Pyear occurs in winter then Pthresh = 2 x Tavg, if 70\% of Pyear occurs in summer then Pthresh = 2 x Tavg + 28, otherwise Pthresh = 2 x Tavg + 14). , Psdry = precipitation of the driest month in summer, Pwdry = precipitation of the driest month in winter, Pswet = precipitation of the wettest month in summer, Pwwet = precipitation of the wettest month in winter. In the northern Hemisphere Summer is defined as AMJJAS and Winter as ONDJFM, the opposite is true for the Southern hemisphere. Due to overlapping criteria, dry (B) climates are prioritised above all others. Temperature is in ˚C and precipitation is cm/month and cm/year.}
\end{table*}

These classifications have three differences to those described by \citep{16}. First, C and D climates follow a 0˚C threshold instead of 3˚C. Secondly, BW and BS are distinguished using a 70\% threshold for precipitation seasonality. Finally, climates C and D subclassifications s and w are made mutually exclusive. In this analysis, each month is set to have the same length of time – one twelfth of a year.

\subsection{Climate Model and Observational Data}
Historical observations of monthly mean temperature and precipitation are from the CRU TS v. 4.05 dataset \citep{17}. Analogous climate model data comes from the ‘historical’ CMIP6 experiments \citep{18}. Six models within the CMIP6 multi-model ensemble were chosen for reasons of data management, these being CanESM5, CanESM5-CanOE, CESM2, CESM2-WACCM, IPSL-CM6A-LR and UKESM1-0-LL.

CMIP6 model data is regridded to 0.5˚ by 0.5˚, the same spatial resolution as CRU TS observations. Antarctica is excluded as observations are limited in this region, and no significant changes in bioclimatic classification are expected in this region.

\subsection{Model performance assessment}
Comparison of KG observed classifications with the CMIP6 models simulated classifications is made for the years 1901-2014. To reduce the effect of short-term variability model and observational data is smoothed with a 30 year centred rolling mean.

The ability of individual models in the CMIP6 ensemble to simulate KG classifications correctly during the historical period is assessed in two ways: (i) Percentage land area that a model has correctly classified for each year relative to observations. (ii) Percentage change in land area classification at each year compared to the initial mean 1901-1931 classifications.

\subsection{Warming Maps}
Future KG classification maps under 1.5, 2 and 4K of annual mean global warming above pre-industrial levels were created from the CMIP6 model ‘ssp585’ 2015-2100 future scenario. We used ssp585 because all models pass 4K under ssp585, which enables us to define changes in bioclimatic zones consistently for these different levels of global warming.

The timing of each warming level is found from the centred 30 year annual mean global surface air temperature above the model’s reference temperature, here defined as 1901 – 1931. Monthly mean anomalies of precipitation and surface air temperature are calculated relative to this same reference period. Where anomaly corrected fields of precipitation and temperature are used, these are calculated as the sum of the mean observations (1901-1931) and an individual model’s anomalies at the specified warming level. Ensemble mean KG classification maps are calculated using the ensemble mean of the anomalous temperature and precipitation fields at each warming level.

\subsubsection{Streamlined Köppen-Geiger classification scheme}
A key goal of bioclimatic classifications is to illustrate climate change in a way that is more intuitive for many people. To this end we designed a  simplified Köppen-Geiger scheme which combines classifications to make changes clearer in both scale and the nature of projected transitions. Additionally, the new scheme implements a more traditional naming system. A breakdown of this streamlined system, and the constituent traditional classifications involved in each of the thirteen streamlined classifications is given in Table 2.

\begin{table}[h]
\begin{tabular}{ll}
 &  \\
 &  \\
\textbf{Streamlined   Classification} & \textbf{Traditional   Classifications} \\
Desert & BWh,   BWk \\
Semi-Arid & BSh,   BSk \\
Tropical   Rainforest & AF \\
Tropical   Monsoon & AM \\
Tropical   Savanna & AW \\
Mediterranean & CSa,   CSb, CSc \\
Subtropical & Cwa,   CWb, CWc, Cfa \\
Oceanic & CFb,   CFc \\
Continental   hot-summer & Dfa,   Dsa, Dwa \\
Continental   cold-summer & DFb,   DSb, DWb \\
Sub   Arctic & DFc,   DFd, DSc, DSd, DWc, DWd \\
Arctic   Tundra & ET \\
Icecap & EF
\end{tabular}
\caption{Breakdown of the streamlined classification scheme and the assignment of traditional classifications within the new scheme.}
\label{tab:my-table}
\end{table}

Difference maps are also plotted to demonstrate the geographical locations of major transitions between bioclimatic classifications. These difference maps plot the ten most significant transitions globally (by total land area).

Classification change matrices are used to quantify bioclimate transitions in terms of global land area, at key levels of global warming. The columns represent the initial classification coverage, and the rows indicate the altered classification distribution. Shading highlights the significance of changes, in terms of the projected change as a fraction of the initial area of a given bioclimatic class.

\section{Results and discussion}
\subsection{Model performance assessment}
To gain insight into the behaviour of individual models,  we create KG maps of individual models and compare them with maps derived from the observed climate. As expected, there is variation in the classification distribution of models and the observational data. For example, desertification in the Amazon is apparent in CanESM5 and CanESM5-CanOE models (Appendix A). This may show that these models have a tendency towards reduced precipitation in the tropics when compared to other models. Another area of disagreement between the models is the change of biome classification in northern Eurasia and America at various levels of global warming. The ensemble mean model state however reduces the effect of individual model discrepancies and compares favourably with observations.
\clearpage

\begin{figure}[h!]
    \includegraphics[width=1\linewidth]{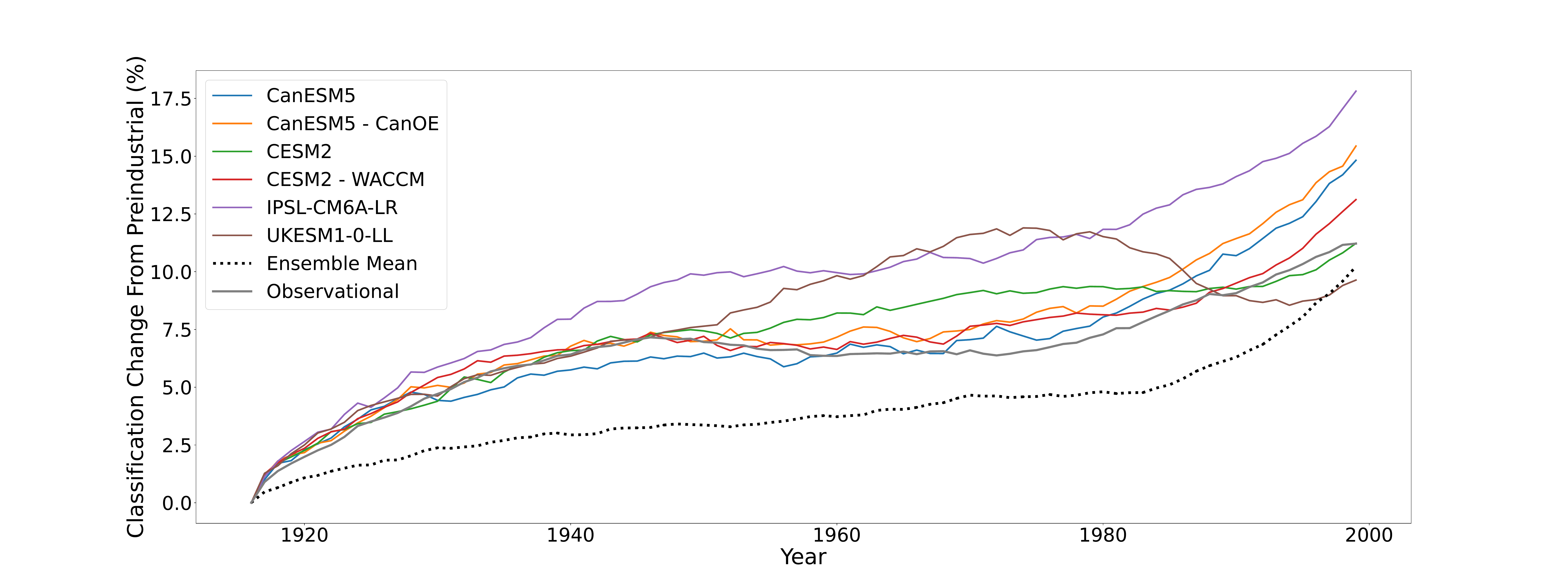}
    \caption{The percentage land area change from the initial 1916 classifications without anomaly correction. Showing the ensemble mean with reduced classification area change compared to models and the observational due to model variation minimisation in the meaning process. Strong agreement between models and the observational.}
\end{figure}

\begin{figure}[h!]
    \includegraphics[width = 17cm]{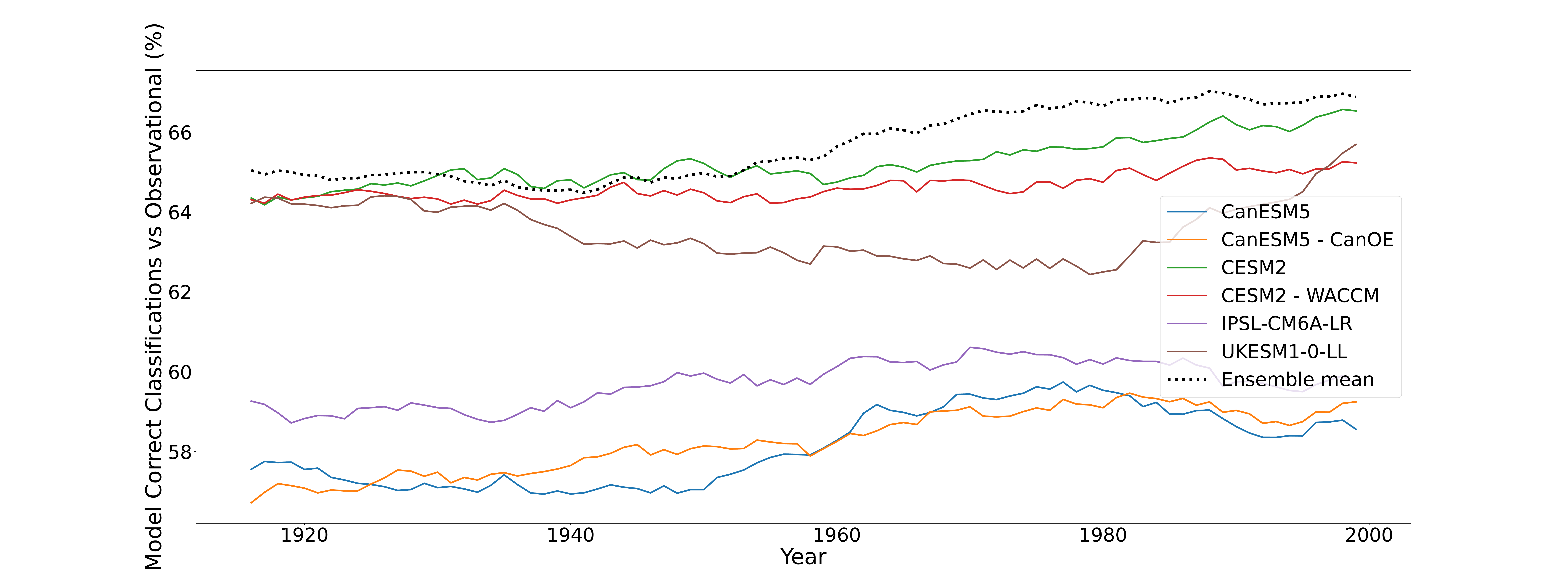}
    \caption{The percentage land area correctly classified without anomaly correction. Showing the ensemble mean as the most consistently correctly classified area compared to the observational due to model variation minimisation in the meaning process. Shows the models initial states separate into two groups.}
\end{figure}
In Figure 1. simulated  classification changes from the CMIP6 historical runs are compared to those calculated from the observed climate. The CMIP6 models broadly capture the degree of expected global classification change. All models show a similar behaviour – a significant change in classifications at the start of the observed period until 1940, the mid-century then presents an approximately constant set of classification with very little change until 1980, where again all models display further changes in climate classification. Although the ensemble mean follows the same pattern as the individual models and the observational data, it shows a lesser degree of change  throughout the observed time period. This reduced variation is inherent to the nature of this ensemble mean; large changes in individual models have their impact reduced in the meaning process. This may lead to the ensemble mean ‘lagging’ the individual models and the observational when creating discrete classes from climate data.

To assess the performance of individual models and their ensemble mean in correctly classifying the bioclimate distribution for a particular year, the percentage land area correctly classified by each model every year is shown in Figure 2.  The results show that the ensemble mean is one of the best performing for classification. This is in contrast to Figure 1 which showed the ensemble mean was one of the worst performing for classification change. The reason is also likely due to the reduced variation in data resulting from the averaging process in the creation of the ensemble mean dataset. The impact of ‘extreme’ values present in each model are averaged out in the ensemble mean provided they are distributed around the `true’ climate values. This would suggest that for individual time points, the ensemble mean is likely to provide the the most reliable projection.
The results from Figure 1 and Figure 2 give insight into the behaviour of ensemble mean datasets and when their application is appropriate. Traditionally the ensemble mean has been taken as the most likely scenario and therefore the most representative of the real-world climate. The results presented here indicate that although the ensemble mean is appropriate for assessing model output at individual points, the ensemble mean does not accurately display the variability evident in real world climate data.
\subsection{Warming maps}

\begin{figure}[h!]
\includegraphics[width = 15cm]{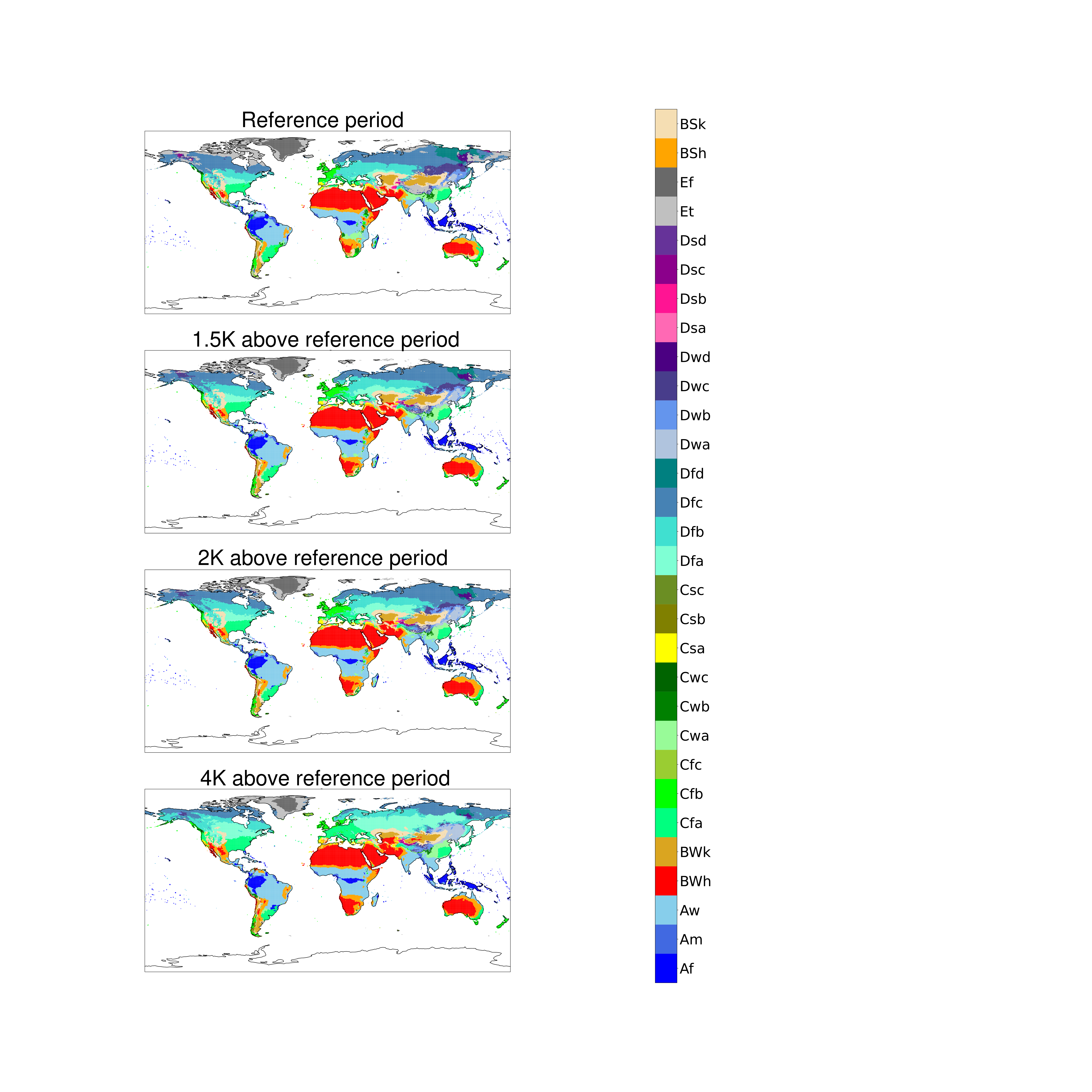}
\caption{Anomaly plot of ensemble mean CMIP6 ssp585 runs for reference period, 1.5K, 2K, and 4K of global warming with the traditional Köppen-Geiger classification system applied. Note the large changes in northern America and Eurasia.}
\end{figure}

Figure 3 shows the ensemble model mean KG classification for 1.5K, 2K and 4K of global warming above the reference period, as well as the no warming classifications. Plots for individual models for the reference period without anomaly correction, and at 1K, 1.5K, 2K, 3K, and 4K of global warming with anomaly correction, are shown in the appendix. Comparison to the reference climate shows that there will be a dramatic changes in bioclimate classification, particularly in the mid- to high latitudes, as the planet warms. These changes become more apparent in Figure 4, which use the streamlined KG classification scheme and are chosen to highlight the ten most significant bioclimatic shifts for each level of warming.

\begin{figure}[h!]
    \includegraphics[width=1\linewidth]{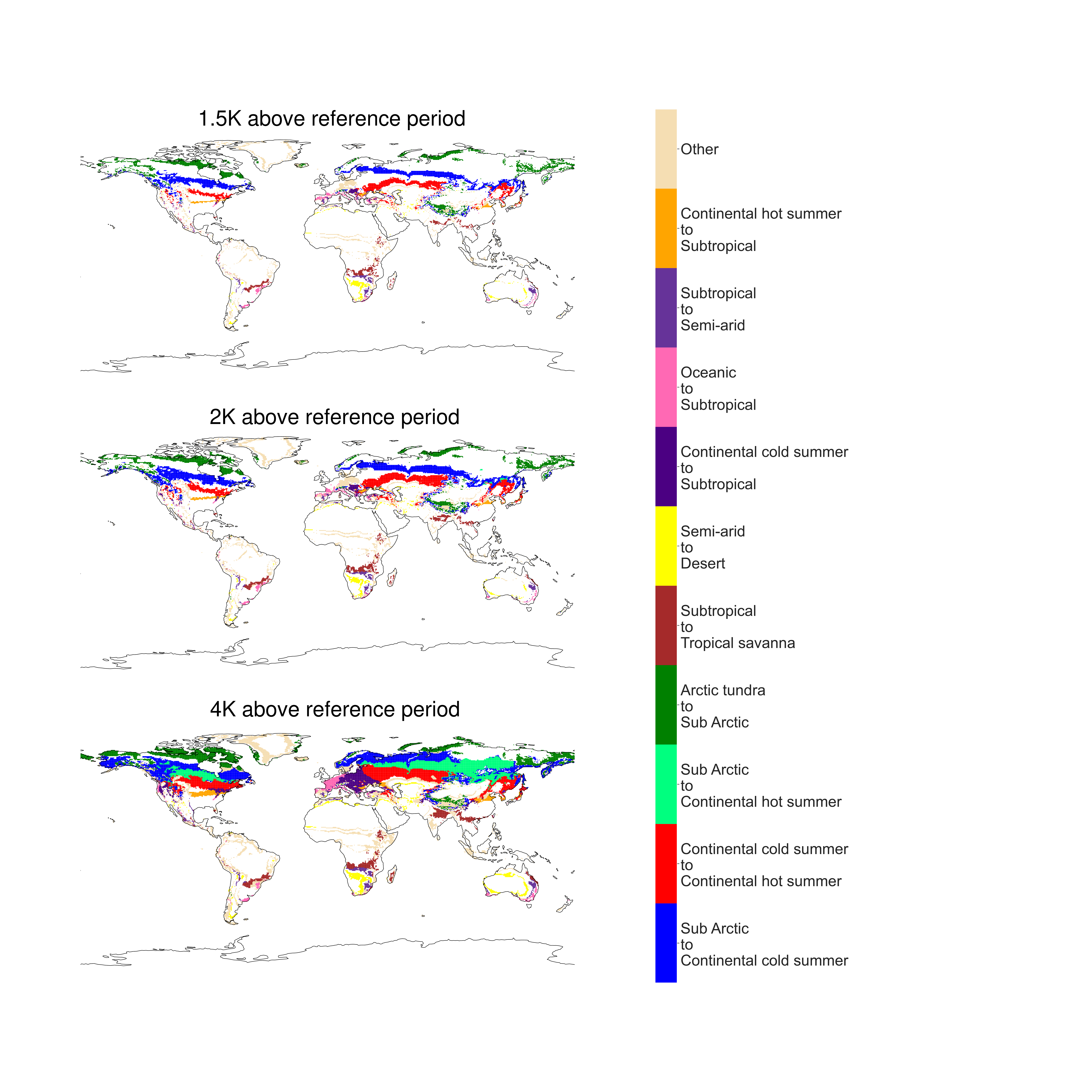}
    \caption{Ensemble mean difference maps highlighting the ten largest classification changes for 4K of warming above the reference period using the streamlined classification system.}
\end{figure}

\begin{figure}[h!]
    \includegraphics[width=13cm]{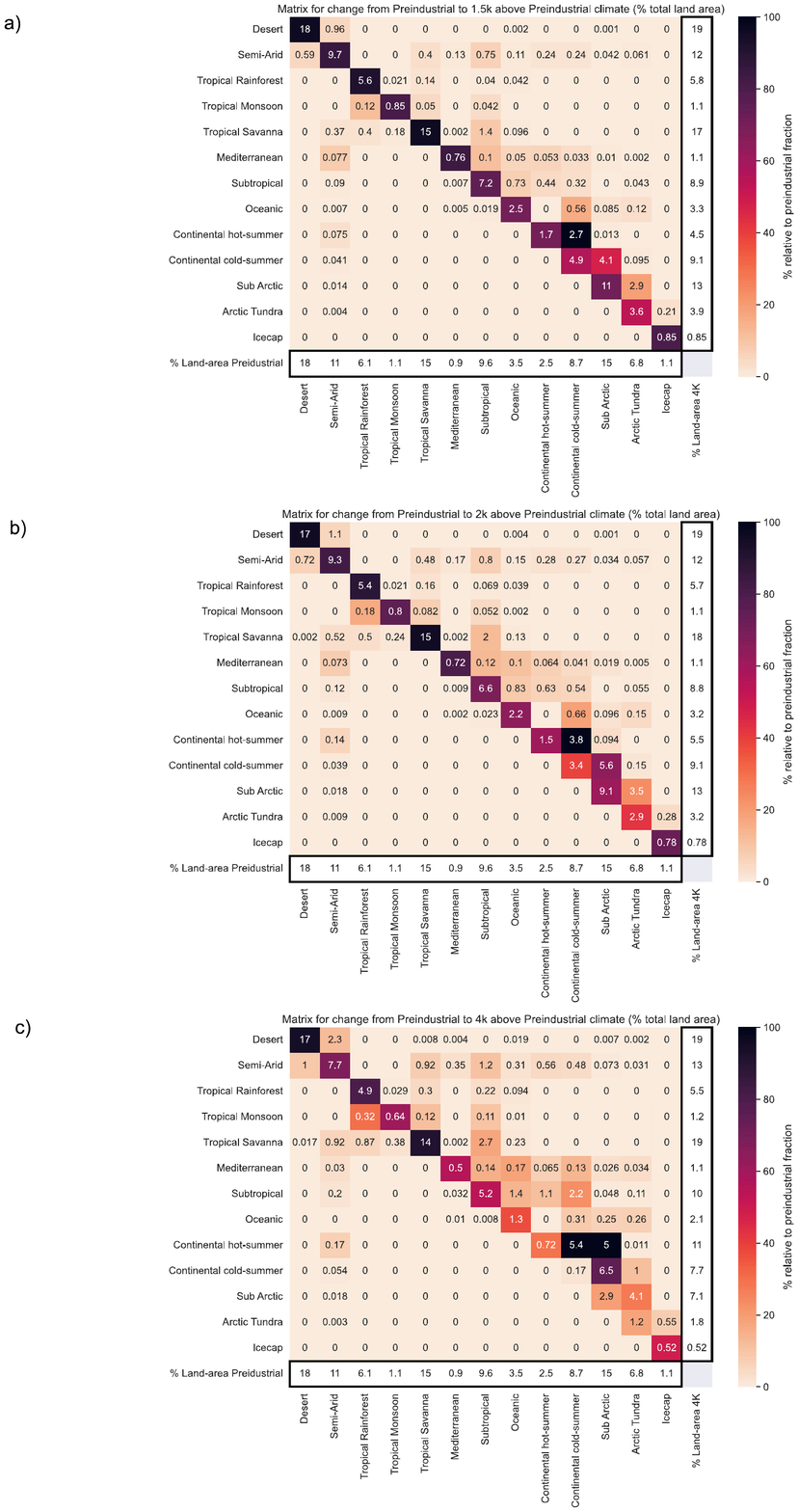}
    \caption{Ensemble mean difference matrices highlighting the  classification changes for levels of warming above the reference period of  a) 1.5K, b) 2K, c) 4K, using the streamlined classification system.}
\end{figure}
\clearpage

These shifts are almost exclusively from wetter and colder classes to drier and hotter ones as the global temperature increases. Large areas undergo desertification in the southern hemisphere. The majority of North America and Northern Eurasia has a shift towards warmer climates as Sub-arctic gives way to continental cold summer, and continental cold summer is replaced by continental warm summer. All changes in classification with the streamlined KG scheme are quantified in Figure 5.

At 4K these areas of classification change represent over 15\% of land area. The change in \% of total land-area in Figure 5a gives
some alarming perspectives, for example, at +4K Arctic Tundra is indicated to cover over 40\% less land-area than in in the reference period. At 2K the models already project significant changes to the global distribution of bioclimates; at 4K these changes become even more pronounced.

\subsection{Sensitivity of bioclimate to global warming}
\begin{figure}[h!]
    \includegraphics[width=1\linewidth]{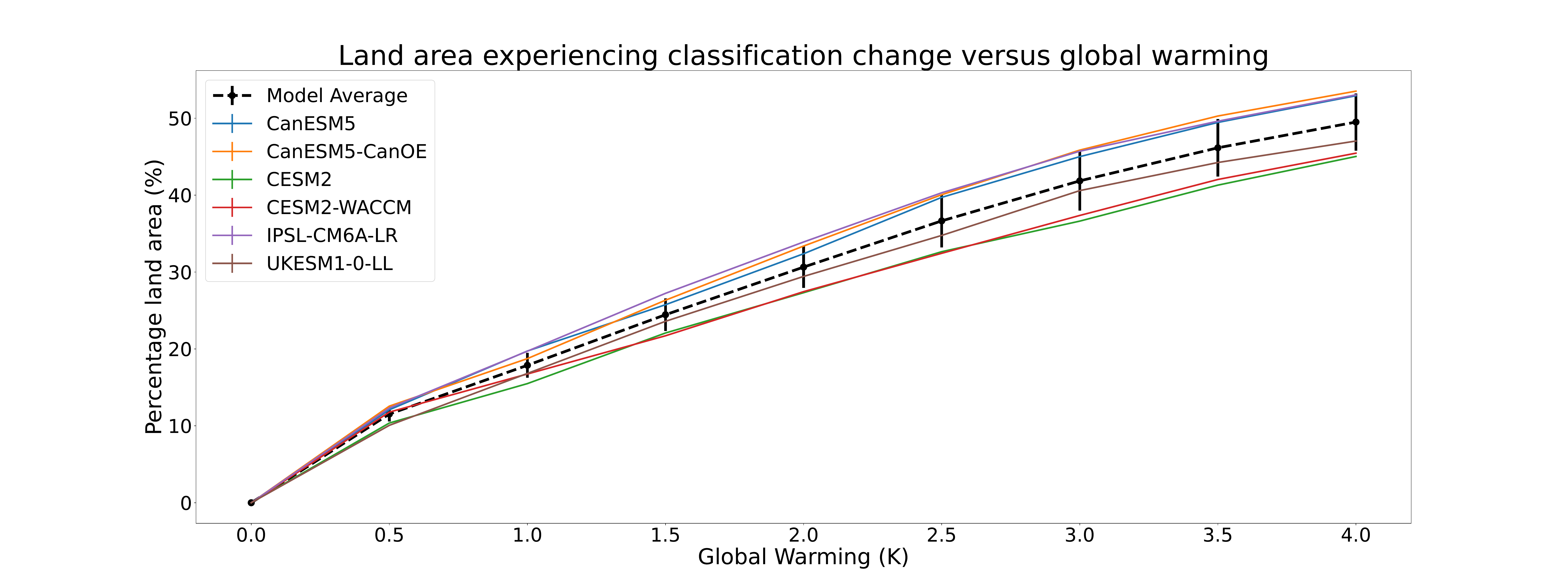}
    \caption{The percentage of land area projected to see a change in bioclimate as a function of global warming, using the traditional Köppen-Geiger classification with anomaly correction. Note the robust agreement between models, which implies an ensemble mean change which is well approximated by:  \begin{math}f = 1-e^{-0.17 \Delta T} \end{math}. }
\end{figure}

Figure 6 displays a weakly-saturating increase with global warming, the fraction of land area that experiences a change in classification follows Eq. (1):
\begin{equation} \label{equation1}
f = 1-e^{-k \Delta T}
\end{equation}
where $f$ is the fraction of land area that experiences a change in bioclimatic classification, $\Delta T$ is the global warming relative to the reference period climate, and $k$ is a fitting parameter. The mean response across the models suggests a value of $k \sim 0.17$ K$^{-1}$.
For the range of global warming of particular interest to the Paris climate agreement (1 to 3 degrees of warming) the land area experiencing a change in bioclimatic classification  is approximately 12\% of the global land per Kelvin of global warming. The total land area (neglecting Antarctica) is approximately 146 million square kilometres, so this implies a bioclimatic change for over 17.5 million square kilometres of land per degree of warming between 1K and 3K. This highlights the benefits of keeping global warming to 1.5K as opposed to 2K of warming, as the 0.5K difference represents an additional bioclimatic change for over 7 million square kilometres of land.
A quantitative distribution of climate classification changes between global warming levels of 1.5K and 2K can be seen in Appendix b (note that this breakdown uses the streamlined KG system and subsequently will not represent all changes included in Fig. 6).

\conclusions  
Despite the difference in climate projections for given greenhouse gas emissions, we present strong evidence that climate models agree well on the extent of bioclimatic change the global land-surface will undergo per degree of global warming. The Köppen-Geiger scheme has been used to present the impact of global warming at 1.5K, 2K, and 4K of warming above reference period levels in the form of climate maps – showing the global distribution of bioclimates, and as graphs and classification change matrices – quantifying the degree of significant climate change for all classifications at various levels of warming.

Despite the fact that these bioclimate classifications are fundamentally climate classifiers, they are designed to represent and correlate with biome distribution. In this way the warming maps and classification changes represent tangible shifts in the global distribution of ecosystems, giving insight into the nature of Earth at various levels of warming. This paper also uses the Köppen-Geiger scheme as a method for climate model verification which is relevant to the impacts of climate change on ecosystems.
The Köppen-Geiger maps at levels of global warming demonstrate the impact that climate change will have. The transition matrices present an easily interpretable method for understanding and quantifying the scale of all classification changes. The results presented by the maps and matrices predict large changes in global bioclimate distribution, with hotter, drier bioclimates expanding and colder, wetter bioclimates shrinking and moving further towards the poles.

The combination of the techniques presented in this paper indicate that the impact of global warming on KG bioclimates is roughly linear for levels of warming between 1 and 3K. We find that 12\% of land will experience a significant change in bioclimate per $^o$C of global warming.

\clearpage


\dataavailability{Based on publicly available CMIP6 data} 




\appendix
\section{}    
\appendixfigures
\begin{figure*}[h]
\includegraphics[width = 16cm]{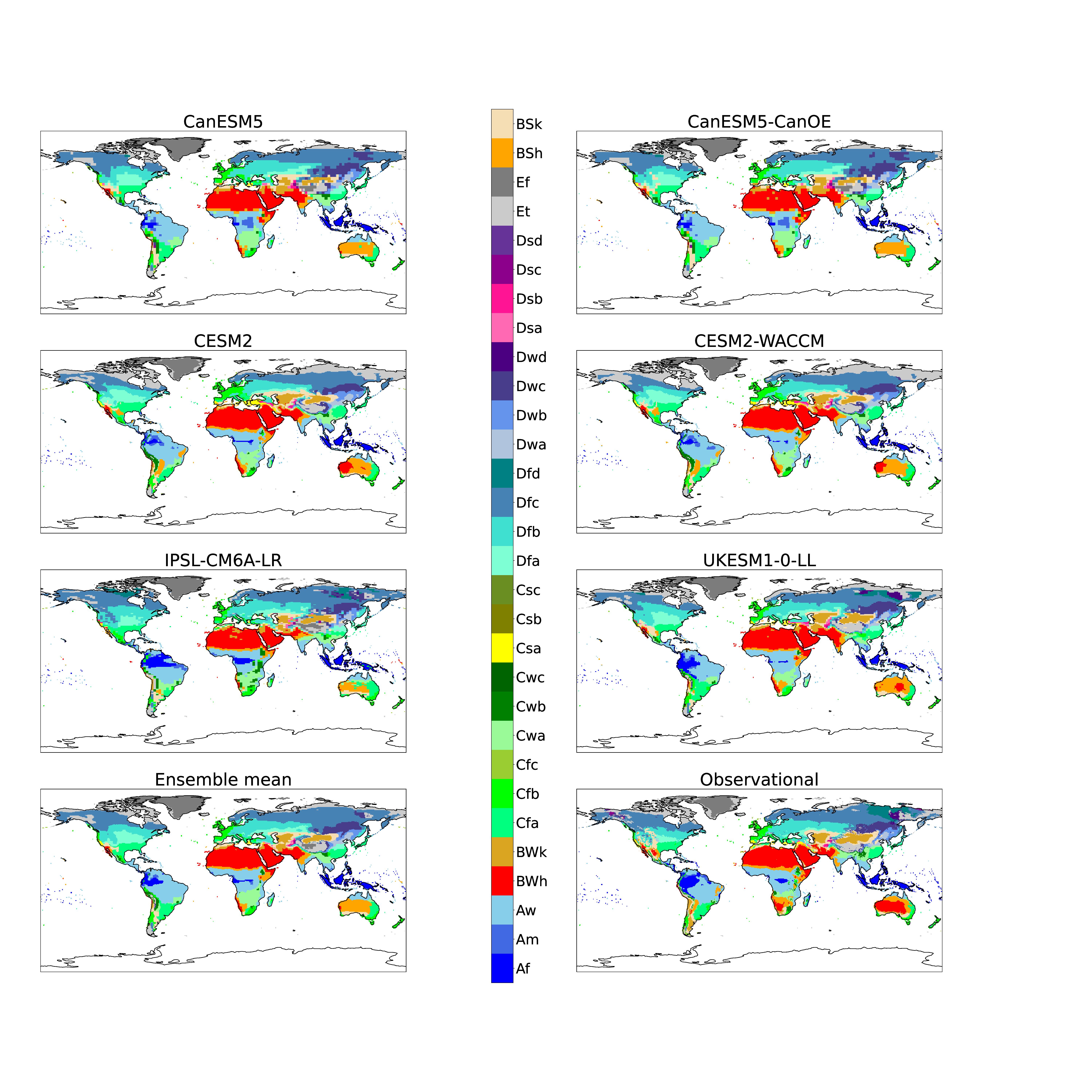}
\caption{All models reference period (1901 - 1931) (No anomaly correction).}
\end{figure*}

\clearpage

\begin{figure*}[t]
\includegraphics[width = 16cm]{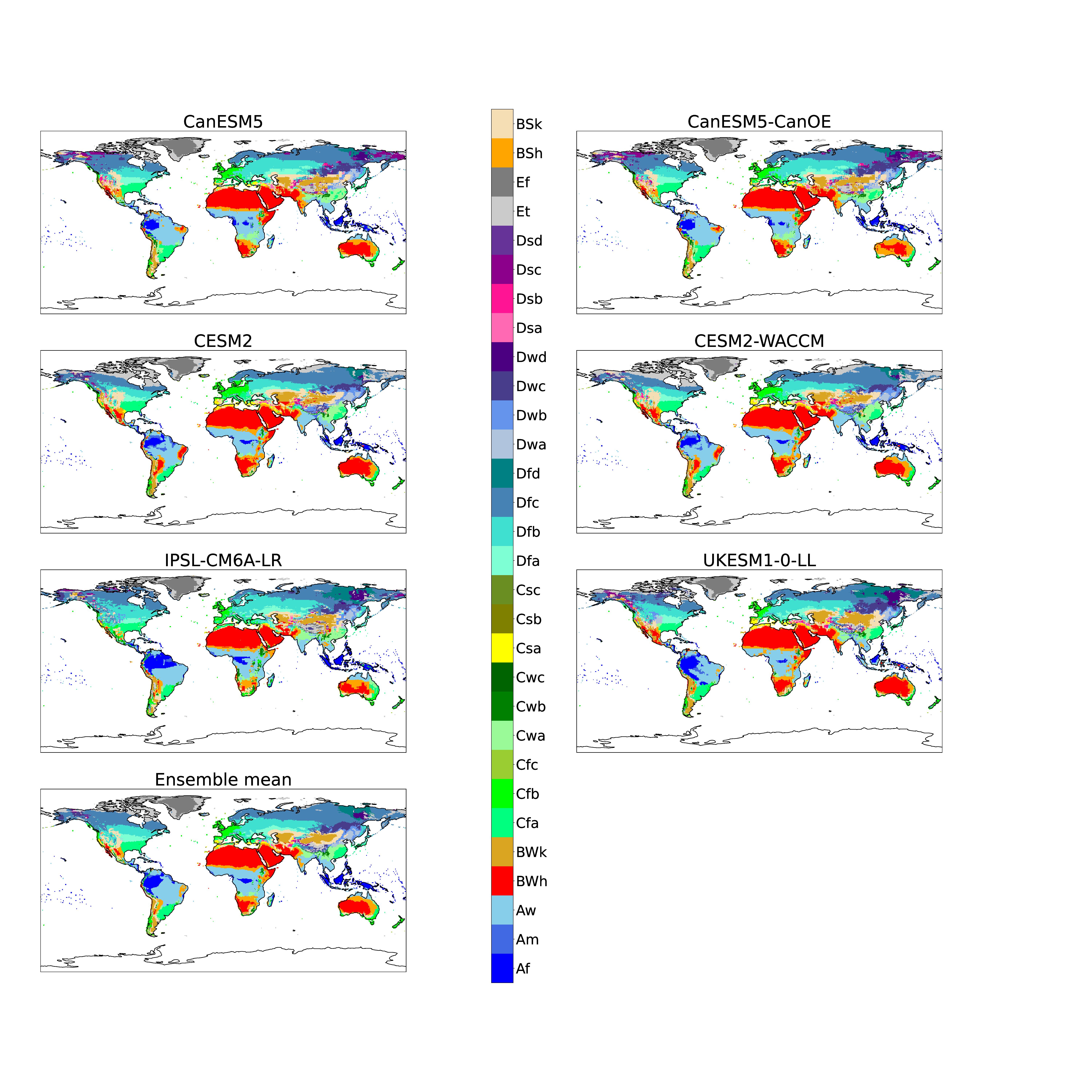}
\caption{All models at +1K with anomaly correction.}
\end{figure*}

\clearpage

\begin{figure*}[t]
\includegraphics[width = 16cm]{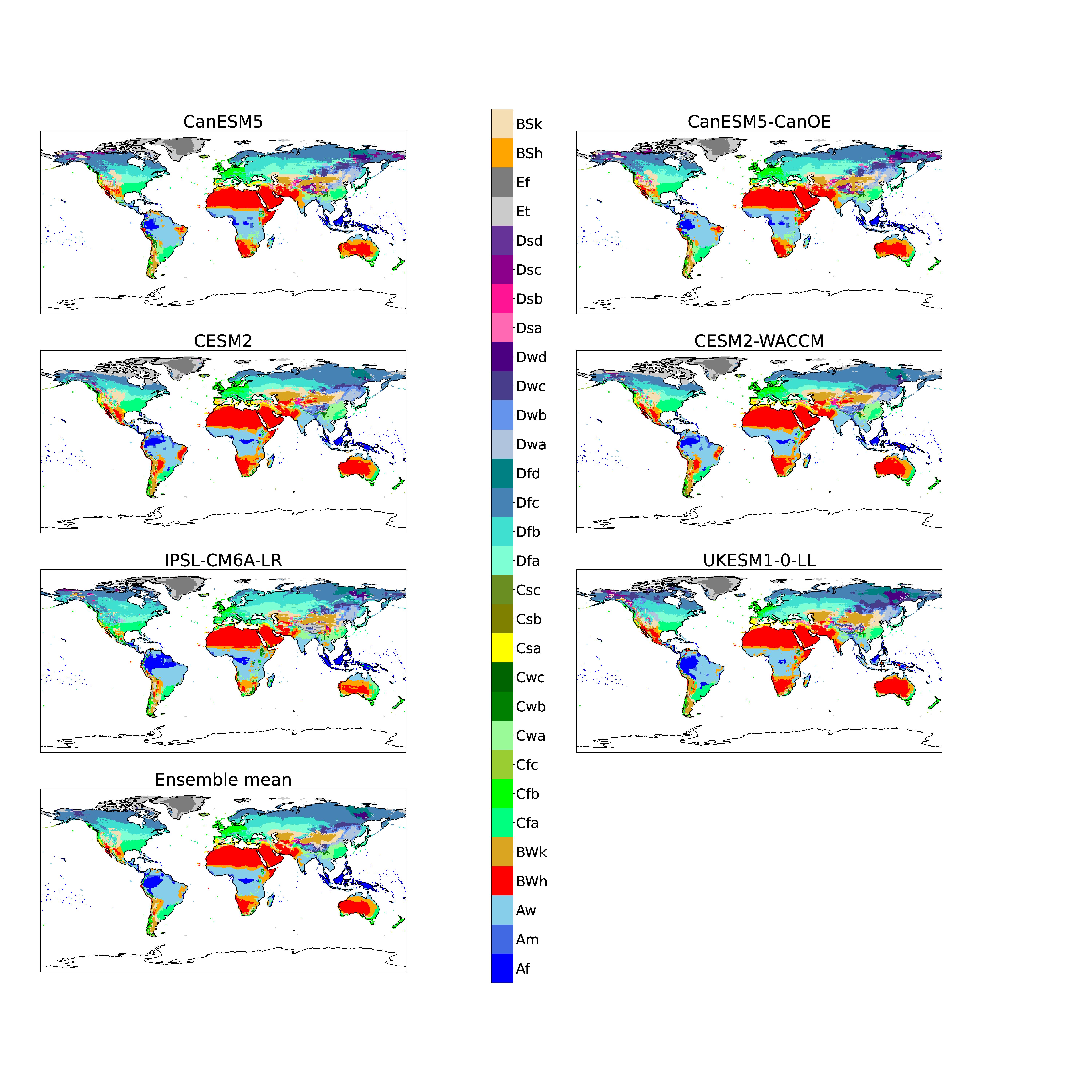}
\caption{All models at +1.5K with anomaly correction.}
\end{figure*}

\clearpage

\begin{figure*}[t]
\includegraphics[width = 16cm]{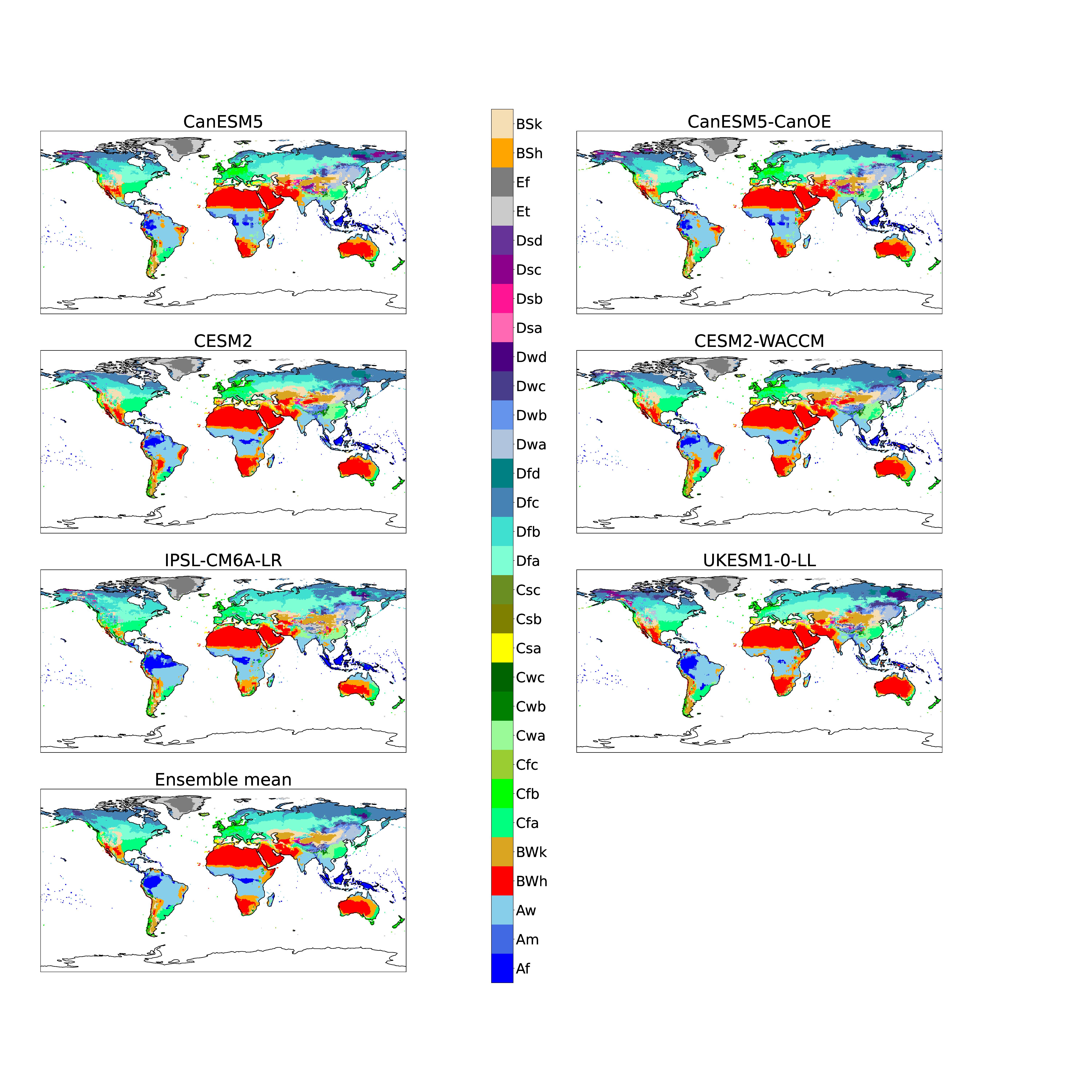}
\caption{All models at +2K with anomaly correction.}
\end{figure*}

\clearpage

\begin{figure*}[t]
\includegraphics[width = 16cm]{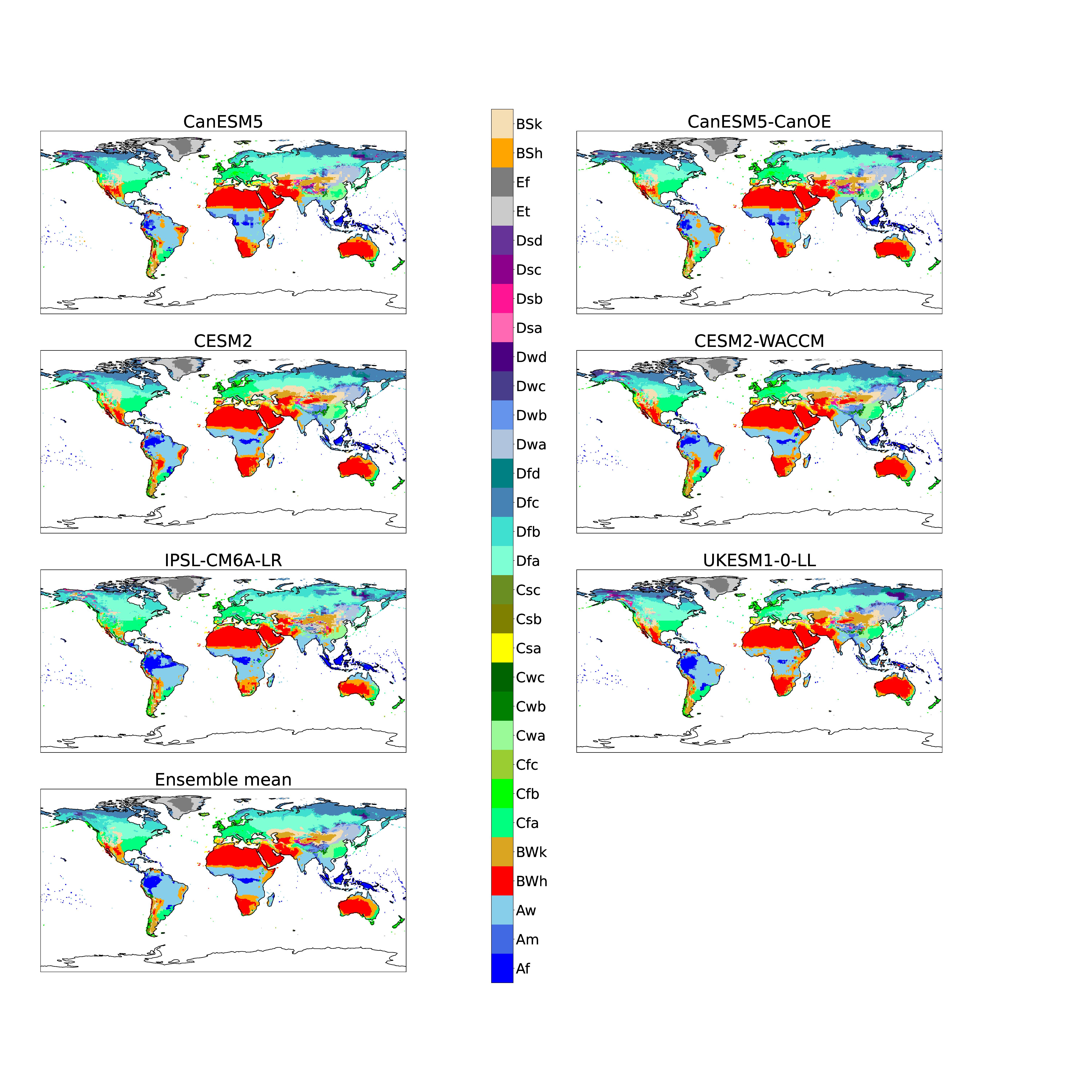}
\caption{All models at +3K with anomaly correction.}
\end{figure*}

\clearpage

\begin{figure*}[t]
\includegraphics[width = 16cm]{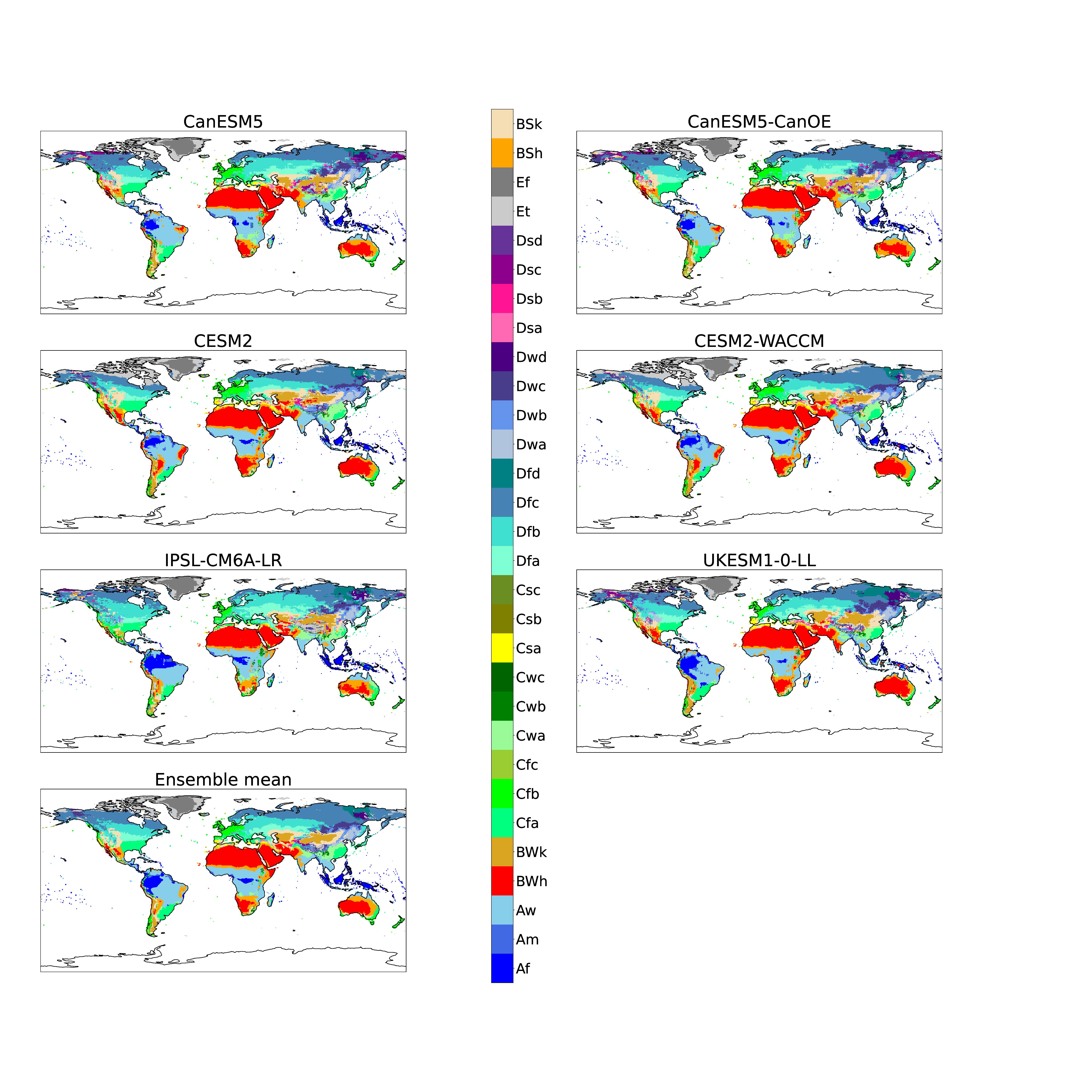}
\caption{All models at +4K with anomaly correction.}
\end{figure*}

\clearpage

\section{}

\begin{figure*}[h]
\includegraphics[width = 16cm]{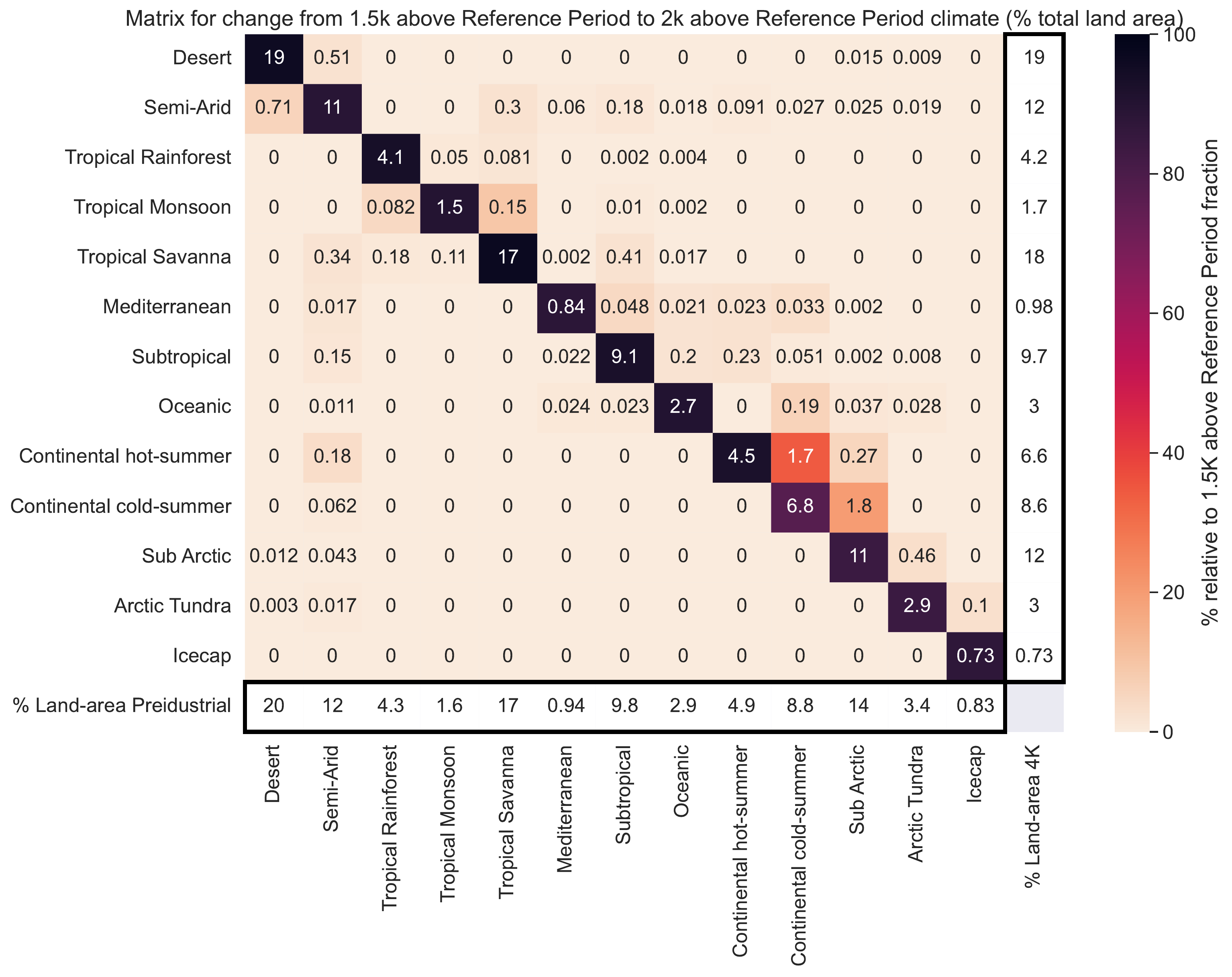}
\caption{Land area bioclimate classification change between 1.5K and 2K of global warming.}
\end{figure*}

\clearpage
\noappendix       







\authorcontribution{MS carried out the data analysis and drafted the paper. MW \& PC advised on the study. All authors contributed to the submitted paper.} 

\competinginterests{No competing interests are present.} 





\bibliographystyle{copernicus}
\bibliography{Ref.bib}

\end{document}